\documentclass[preprint,showpacs,preprintnumbers,amsmath,amssymb]{revtex4}

\usepackage{graphicx}
\usepackage{dcolumn}
\usepackage{bm}

\begin{document}

\title{Complete population transfer in a degenerate 3-level atom}

\author{Kh.Yu. Rakhimov\footnote{Permanent address: 
Department of Heat Physics, Uzbekistan Academy of Sciences, 
28 Katartal St., Tashkent 700135, Uzbekistan}}
\author{J.H. McGuire}
\author{Kh.Kh. Shakov}
\affiliation{Physics Department, Tulane University, New Orleans, LA 70118 USA.}

\date{\today}

\begin{abstract}
We find conditions required to achieve complete population transfer,
via coherent population trapping, from an initial state to a designated 
final state at a designated time in a degenerate 3-level atom, where 
transitions are caused by an external interaction. 
Complete population transfer from an initially occupied state 1
to a designated state 2 occurs under two conditions.
First, there is a constraint on the ratios of the transition matrix elements
of the external interaction.
Second, there is a constraint on the action integral over the interaction,
or "area", corresponding to the phase shift induced by the external interaction.
Both conditions may be expressed in terms of simple odd integers. 
\end{abstract}

\pacs{32.80.-t,42.50.-p,32.80.Qk}

\maketitle

\section{Introduction}

Population control in quantum systems, namely transfer of electrons
from an ensemble of atoms all in the same initial state to specified
final states, is used in problems ranging from coherent population trapping 
\cite{alz76,asp88,holland95,scully02,metcalf02}, 
including electromagnetically induced transparency \cite{bol91,harris97,Scully,lukin01}, 
and quantum computing 
\cite{cirac97,Nielsen,Bouwmeester,mair02,haroche01,knight02,rgb02},   
to chemical dynamics \cite{Rabitz93,Shapiro02}.    
These problems are modeled in terms of an $n$-level (often $3$-level) atom 
interacting with a strong external field 
\cite{Shore,Milonni88,Meystre,Eberly75,Guerin97,bbes77}. 
In this paper we consider population control in a nearly degenerate 
$3$-level atom.  We show that the $3$-level atom is relatively easy to understand 
in the degenerate limit, where all $3$ states have the same energy.
We specifically show how to achieve complete population transfer in a degenerate 
$3$-level atom where the matrix elements of the external interaction have a common time dependence.

Our degenerate approximation \cite{sm02} is mathematically similar to the 
rotating wave approximation (RWA) \cite{Shore,Eberly75}, which
has been widely applied to both $2$ and $3$-level atomic models.
However, in the RWA degenerate atomic states are not used.
Instead one tunes the frequency of the external field 
to the frequency difference of two non-degenerate levels
so that the detuning parameter \cite{Milonni88} 
tends to zero.  Thus, in RWA the initial state of an atom plus one
photon is degenerate in energy with the final state of the atom.
One advantage of using degenerate atomic states, as done in this paper, 
is that one may use external interactions with a broad range of frequencies.   
Another advantage is that the interaction frequency 
can be used as a control parameter, e.g. to vary the duration of time
that the transferred population remains in the designated
state, or to reduce the population leakage that occurs
when the three levels are not fully degenerate.

In the next section we derive analytic formulae for
the probabilities, $P_{1,2,3}(t)$, that the electron is in level 1, 2 or 3 at time $t$.
Then we seek conditions for complete population transfer
to a designated level that is initially unoccupied at a designated time $t_0$.  
This places conditions on both the relative strengths 
of the interaction matrix elements, $V_{ij}(t)$,  
and the action integral, $A(t_0) = \int_0^{t_0} V(t') dt'$.
We show that these conditions may be expressed in terms of two odd integers,
$n_1$ and $n_2$.  Some analysis is done for population leakage,
when the levels are not quite degenerate.
Calculations are presented.

\section{Theory}

Let us consider an $n$-level atom interacting with an external field, $V_{ext}(\vec{r},t)$.
The total Hamiltonian for this system is $H = H_0 + V_{ext}(\vec{r},t)$.  The $n$ eigenstates, 
$\phi_k$, and corresponding eigenenergies, $E_k$, of $H_0$ are assumed to be known.
The total wavefunction may be expanded in terms of the known eigenstates, namely, 
$\Psi(t) = a_1(t) \phi_1 + a_2(t) \phi_2 + \cdots + a_n(t) \phi_n$.
With atomic units, using $i \dot{\Psi} = (H_0 + V_{ext}(\vec{r},t)) \Psi$, 
with $H_0 \phi_k  = E_k \phi_k$ and $\int \phi^*_j \phi_k d\vec{r} = \delta_{jk}$,
one then obtains  \cite{Milonni88},  
\begin{eqnarray}
\label{nexact}
i \dot{a}_j(t) = E_j a_j(t) + \sum_{k=1}^n V_{jk}(t) a_k(t)  \ \ ,
\end{eqnarray}
where $V_{jk}(t) = \int \phi^*_j V_{ext}(\vec{r},t) \phi_k d \vec{r}$. 
These equations are exact for an $n$-level atom.

We now require that the system be degenerate, namely that
all the energies, $E_j$, are the same. 
Since the zero point of energy is arbitrary, one may generally set $E_j = 0$.
These conditions give the coupled equations for a degenerate $n$-level system, namely,
\begin{eqnarray}
\label{ndegen}
i \dot{a}_j(t) =  \sum_{k =1}^n V_{jk}(t) a_k(t)  \ \ .
\end{eqnarray}
We use the initial condition $a_1(0) = 1$, and $a_j(0) = 0$ for $j \neq 1$.
We additionally require that all of the $V_{jk}(t)$ have the same time dependence.
Here we also take $V_{jk} = V_{kj}$ to be real.

\subsection{Degenerate 3-level atom}

The degenerate 3-level atom can be solved analytically.  
Recalling that the $V_{jk}(t)$ have a common time dependence,
we choose $V(t) = V_{23}(t) = \alpha^{-1} V_{12}(t) = \beta^{-1} V_{13}(t) 
= \epsilon_j^{-1} V_{jj}(t)$.  
A relatively simple solution, presented next, is found by taking $\epsilon_j = 0$.
More general solutions for $\epsilon_j \neq 0$ are discussed in an appendix.
With $\epsilon_j = 0$ and $n = 3$, Eq(\ref{ndegen}) becomes,
\begin{eqnarray}
\label{3degen}
i \dot{a}_1(t) &=&   \alpha V(t) a_2(t) + \beta V(t) a_3(t)  \ \ , \\ \nonumber
i \dot{a}_2(t) &=&  \alpha V(t) a_1(t) +  V(t) a_3(t)  \ \ ,  \\ \nonumber
i \dot{a}_3(t) &=&  \beta V(t) a_1(t)  + V(t) a_2(t)  \ \ .
\end{eqnarray}

Now consider the linear combination $c(t) =  a_1(t) + x \ a_2(t) + y \ a_3(t)$,
where $x$ and $y$ are some time-independent coefficients that we next determine.
Then,
\begin{eqnarray}
\label{3trial}
i \dot{c}(t) &=& i \dot{a}_1(t) + x \ i \dot{a}_2(t) + y \ i \dot{a}_3(t) \\ \nonumber
 &=&  \   \alpha V(t) \ a_2(t) + \beta V(t) \ a_3(t) 
	 + x  \ [\alpha V(t) \ a_1(t)  + V(t) \ a_3(t)]   \\ \nonumber  
	&& + y \ [\beta V(t) \ a_1(t) + V(t) \ a_2(t) ]  \ \ .
\end{eqnarray}
Set $z = \alpha x + \beta y$.  Then,
$i \dot{c}(t) = z V(t)[ a_1(t) +   
 (\frac{\alpha +  y}{\alpha x + \beta y}) a_2(t)
+ (\frac{\beta + x }{\alpha x + \beta y}) a_3(t) ]$.
We require that $ i \dot{c}(t) =  z V(t) c(t)$.
This holds if and only if $ x = (\alpha +  y)/(\alpha x + \beta y)$ 
and $y = (\beta + x )/(\alpha x + \beta y)$.
After some algebra this leads to the useful cubic equation,
\begin{eqnarray}
\label{cubicy}  
 (\beta^2 -\alpha^2)y^3  +\alpha(2-\alpha^2-\beta^2)y^2 
   +(2\alpha^2-\beta^2-1)y  + \alpha(\beta^2-1) = 0  \ \ .
\end{eqnarray}
This determines three sets of eigenvalues, $\{x_j\}$, $\{y_j\}$ and $\{z_j\}$,  
and three eigenfunctions, $ c_j(t) = e^{-i z_j  A(t)}$, where 
$A(t) = \int_0^t V(t')dt'$.  Specifically,
$c_j=\sum_{i=1}^3{\cal M}_{ji}a_i$, where,
\begin{equation}
{\cal M}=\left(\begin{array}{ccc}
1&x_1&y_1\\
1&x_2&y_2\\
1&x_3&y_3\\
\end{array}\right)
\end{equation}
This matrix, ${\cal M}$, may be inverted, namely,
\begin{equation}
\label{Minv}
{\cal M}^{-1}=\frac{1}{\Delta}\left(\begin{array}{ccc}
x_2y_3-x_3y_2&x_3y_1-x_1y_3&x_1y_2-x_2y_1\\
y_2-y_3&y_3-y_1&y_1-y_2\\
x_3-x_2&x_1-x_3&x_2-x_1\\
\end{array}\right)
\end{equation}
where
  $ \Delta=\det({\cal M})=x_1y_2+x_2y_3+x_3y_1-x_1y_3-x_2y_1-x_3y_2 $.

Now one may express the unperturbed state amplitudes, $a_j(t)$,
in terms of the dressed-state amplitudes, $c_j(t)$.  Using
$c_j(t) = e^{-iz_jA(t)}$, one has  
$a_i(t) = \sum_{j=1}^3 {\cal M}^{-1}_{ij} c_j(t) 
= \sum_{j=1}^3 {\cal M}^{-1}_{ij} e^{-iz_jA(t)}$. 
This leads to, 
\begin{eqnarray}
\label{Pj}
P_1(t)&=&|a_1(t)|^2=\frac{1}{\Delta^2}[(x_2y_3-x_3y_2)^2+(x_3y_1-x_1y_3)^2
	+(x_1y_2-x_2y_1)^2 \nonumber \\
&&+2(x_2y_3-x_3y_2)(x_3y_1-x_1y_3)\cos((z_1-z_2)A(t)) \nonumber \\
&&+2(x_2y_3-x_3y_2)(x_1y_2-x_2y_1)\cos((z_1-z_3)A(t))\nonumber \\
&&+2(x_3y_1-x_1y_3)(x_1y_2-x_2y_1)\cos((z_2-z_3)A(t))] \ \ ,  \nonumber \\
P_2(t)&=&|a_2(t)|^2=
\frac{1}{\Delta^2}[(y_2-y_3)^2+(y_3-y_1)^2+(y_1-y_2)^2\nonumber \\
&&+2(y_2-y_3)(y_3-y_1)\cos((z_1-z_2)A(t))\nonumber \\
&&+2(y_2-y_3)(y_1-y_2)\cos((z_1-z_3)A(t))\nonumber \\
&&+2(y_3-y_1)(y_1-y_2)\cos((z_2-z_3)A(t))] \ \ ,  \nonumber \\
P_3(t)&=&|a_3(t)|^2=
\frac{1}{\Delta^2}
[(x_3-x_2)^2+(x_1-x_3)^2+(x_2-x_1)^2\nonumber \\
&&+2(x_3-x_2)(x_1-x_3)\cos((z_1-z_2)A(t))\nonumber \\
&&+2(x_3-x_2)(x_2-x_1)\cos((z_1-z_3)A(t))\nonumber \\
&&+2(x_1-x_3)(x_2-x_1)\cos((z_2-z_3)A(t))]  \ .
\end{eqnarray}
Eq(\ref{Pj}) generally gives the transition probabilities to all three final states
for arbitrary $V_{jk}(t)$, and corresponds to the more general expression \cite{msr03}, 
$P_k(t) 
= \sum_{i}^n \ \sum_{j}^n  \  
{\cal M}_{ki}^{-1}{\cal M}_{kj}^{-1}  \cos[(z_i - z_j) A(t)] $.
Since the $x_{j}$'s, $y_j$'s and $z_j$'s vary with the $V_{jk}(t)$,
one may then seek conditions on the matrix elements $V_{jk}$ such that the
electron populations $P_k(t_0)$ take any desired values at any specified time, $t = t_0$.

\subsection{Complete population transfer}

Next we seek conditions on the $V_{jk}(t)$ such that at time $t_0$ the electron 
population is completely transferred from its initial state $i = 1$ to a different
final state, $i = 2$ or $3$.  
As shown from Eq(\ref{ab}) in an appendix, the condition that 
$\epsilon_j = 0$ yields $\alpha = 2/y -y$ and $\beta = \pm1$.
From the appendix taking $\beta = +1$ in case i) and 
using $r = \pm \sqrt{\frac{2}{n_1 n_2}}$, one has,
\begin{eqnarray}
\label{xyz}
\{x_1, x_2, x_3\} &=& \{ 1, 1, -1 \}  \ \ ,  \\ \nonumber
\{y_1, y_2, y_3\} &=& \{ y_+, y_-, 0 \} = \frac{1}{2}\{ -\alpha + \sqrt{\alpha^2 + 8},  
	 \ -\alpha - \sqrt{\alpha^2 + 8}, \ 0 \}  \\ \nonumber
 &=&  r \ \{ n_1, \ -n_2, \ 0 \} \ \ , \\ \nonumber
\{z_1, z_2, z_3\} &=& \{ \alpha + y_+, \alpha + y_-, -\alpha \} 
	=  \frac{1}{2}\{ \alpha + \sqrt{\alpha^2 + 8},
         \ \alpha - \sqrt{\alpha^2 + 8}, \  -2\alpha \} \\ \nonumber 
	&=& \{ -y_-, -y_+, -\alpha \} 
	 = r \ \{ n_2, \  - n_1, \ n_1 - n_2 \} \ \ ,  \\ \nonumber
\Delta &=& 2(y_- - y_+) = -2 \sqrt{\alpha^2+8} = -2 r (n_1 + n_2) \ \ . 
\end{eqnarray} 
Here $n_1$ and $n_2$ are integers whose values are specified below. 
This leads to certain allowed values of the action integral $A(t_0)$ and the relative
interaction strength, $\alpha$, namely,
\begin{eqnarray}
\label{qnumbers'}
    3 r \  A(t_0)&=& \pm 3 \sqrt{\frac{2}{n_1n_2}} \ \ A(t_0) = \pi  \ \ , \nonumber \\ 
 \alpha &=& V_{12}(t)/V_{23}(t) =  r \ (n_2 - n_1)  \ \ , \nonumber \\ 
 \beta  &=& V_{13}(t)/V_{23}(t) = \pm 1  \ \ . 
\end{eqnarray}
The same allowed values of $A(t_0)$, $\alpha$ and $\beta$ are found in all cases, 
shown in the appendix. 

One may use the values of the $x_j$, $y_j$, $z_j$ and $\Delta$ from
Eq(\ref{xyz}) in Eq(\ref{Pj}) to obtain explicit 
expressions for the transition probabilities, namely,
\begin{eqnarray}
\label{Pj'}
	P_1(t) &=& \frac{1}{2 (n_1 + n_2)^2} \{ n_1^2 + n_2^2 
	      +  n_1 n_2 [ 1  +  \cos ((n_1 + n_2) r A(t))] \nonumber \\
		&& + (n_1 + n_2) [n_1 \cos ((2 n_1 - n_2) r A(t))
			+ n_2 \cos ((2n_2 - n_1) r A(t)) ] \}  \ \ , \nonumber \\
	P_2(t) &=& \frac{1}{2 (n_1 + n_2)^2} \{ n_1^2 + n_2^2 
	      +  n_1 n_2 [ 1  +  \cos ((n_1 + n_2) r A(t))] \nonumber \\
		&& - (n_1 + n_2) [ n_1 \cos ((2 n_1 - n_2) r A(t))
			+ n_2 \cos ((2n_2 - n_1) r A(t)) ] \}  \ \ , \nonumber \\
	P_3(t) &=& \frac{2 n_1 n_2}{(n_1 + n_2)^2}   \ \sin^2{[ \ (\frac{n_1 + n_2}{2})
		 \ r A(t) \ ] } \ \ , 
\end{eqnarray}
where 
$n_1$, $n_2$ are odd integers. 
The constraints on the values of $n_1$ and $n_2$ are as follows.
The condition $P_1 = 0$ requires that 
$\frac{1}{3}(2n_1 - n_2)$ and $\frac{1}{3}(2n_2 - n_1)$ are both odd integers
(i.e. $n_0$ and $n_o'$),
and also requires that $\frac{1}{3}(n_1 + n_2)$ is an even integer.
The condition $P_2 = 1$ imposes the same requirements.
The condition $P_3 = 0$ only requires that $\frac{1}{3}(n_1 + n_2)$ be an even integer,
but $P_3 \geq 0$ requires that $n_1 n_2 \geq 0$.

Finally a slightly neater solution may now be obtained from the requirements 
above that $2n_1 - n_2 = 3n_o$ and $2n_2 - n_1 = 3n_o'$, where $n_o$ and $n_o'$
are both odd integers (either positive or negative).  It is easily shown that
$n_1 + n_2 = 3(n_o + n_o')$ and $n_1 n_2 = 2n_o^2 + 5 n_o n_o' + 2n_o^2$.
Then $r = \pm(n_o^2 + \frac{5}{2} n_o n_o' + n_o^{\prime 2})^{-\frac{1}{2}}$, 
$\alpha =  r (n_o' - n_o)$ and $3r A(t_0)= \pi$. Now,
\begin{eqnarray}
\label{Pj''}
	P_1(t) &=& \frac{1}{2 (n_1 + n_2)^2} \{ n_1^2 + n_2^2 
	      +  n_1 n_2 [ 1  +  \cos ( \ (n_o + n_o') 
		\pi \left( A(t)/A(t_0) \right) \ )] \nonumber \\
		&& + (n_1 + n_2) 
		[n_1 \cos ( \ n_o  \pi \left( A(t)/A(t_0) \right) \ )  
   + n_2 \cos ( \ n_o' \pi \left( A(t)/A(t_0) \right) \ )] \}  \ \ , \nonumber \\
	P_2(t) &=& \frac{1}{2 (n_1 + n_2)^2} \{ n_1^2 + n_2^2 
	      +  n_1 n_2 [ 1  +  \cos ( \ (n_o + n_o') 
		\pi \left( A(t)/A(t_0) \right) \ )] \nonumber \\
		&& - (n_1 + n_2) [n_1 \cos ( \ n_o  
		\pi \left( A(t)/A(t_0) \right) \ )  
   + n_2 \cos ( \ n_o' \pi \left( A(t)/A(t_0) \right) \ )] \} \ \ ,  \nonumber \\
	P_3(t) &=& \frac{2 n_1 n_2}{(n_1 + n_2)^2} \ \sin^2( \ \frac{1}{2}(n_o + n_o')
		 \pi \left( A(t)/A(t_0) \right) \ ) \ \ . 
\end{eqnarray}
Here $n_o$ and $n_o'$ are arbitrary odd integers, 
$n_1 = 2 n_o + n_o'$, and $n_2 = n_o + 2 n_o'$.
At $t = t_0$,  $ A(t)/A(t_0) = 1$, so that $P_2(t_0) = 1$
with $P_1(t_0) = P_3(t_0) = 0$.  This yields complete population
transfer from level 1 to level 2 at $t = t_0$.  
We regard $n_o$ and $n_o'$ as the more fundamental numbers 
since they obey the simplest rules.

\section{Results}

Some allowed values of the action integral, $A(t_0)$, and the relative
interaction strength, $\alpha$, are given in table I.
This table includes values of $\{n_1,n_2\}$, $\{n_o,n_o'\}$ and 
all three $\{k,k'\}$ cases defined in the appendix. 
From more complete numerical output we confirm that $n_1$ and $n_2$ each
acquire all possible odd integer values, although values of
the product $n_1 \cdot n_2$ are restricted.  
This is consistent with the condition that 
$n_1 = 2 n_o + n_o'$, and $n_2 = n_o + 2 n_o'$,
where $n_o$ and $n_o'$ are arbitrary odd integers.
It is also evident that the even integer, $n_e = n_o + n_o'$, 
takes on all even values.  
One may also show algebraically that for each value of
the even integer $k$ in case i) there are two odd values of
an odd integer $k$ in case ii), and vice versa.
The sets of integers $\{n_1,n_2\}$, $\{n_o,n_o'\}$ and $\{k,k'\}$ are redundant.
The three sets of $\{k,k'\}$ correspond to a single set $\{n_1,n_2\}$,
while the $\{n_o,n_o'\}$ are in one to one correspondence with the $\{n_1,n_2\}$.

\begin{table}
\centering
\label{tab:1}
\begin{tabular}{||c|cc|cc|cc|cc|c|c||}
\hline
\multicolumn{1}{||c|}{$n_1\cdot n_2$} &\multicolumn{1}{c}{$n_1$} &\multicolumn{1}{c|}{$n_2$}
&\multicolumn{1}{c}{$n_e$} &\multicolumn{1}{c|}{$n_o$}
&\multicolumn{1}{c}{$n_o$} &\multicolumn{1}{c|}{$n_o'$}
&\multicolumn{1}{c}{$n_o'$} &\multicolumn{1}{c|}{$n_e$}
&\multicolumn{1}{c|}{$A(t_0)$} &\multicolumn{1}{c||}{$\alpha$} \\
& & & $k$ & $-k'$ & $k$ & $k'$ & $-k$ & $k'$ & & \\
\hline
5 \ \ \  &$\pm 1$  &$\pm 5$  &$\pm 2$ &$\mp  1$ &$\mp  1$ &$\pm  3$ &$\pm  3$ &$\pm 2$ &$\pm 1.656$ &$\mp 2.530$ \\
5 \ \ \  &$\pm 5$  &$\pm 1$  &$\pm 2$ &$\pm  3$ &$\pm  3$ &$\mp  1$ &$\mp  1$ &$\pm 2$ &$\pm 1.656$ &$\pm 2.530$ \\
9 \ \ \  &$\pm 3$  &$\pm 3$  &$\pm 2$ &$\pm  1$ &$\pm  1$ &$\pm  1$ &$\pm  1$ &$\pm 2$ &$\pm 2.221$ &$    0.000$ \\
11\ \ \  &$\pm 1$  &$\pm 11$ &$\pm 4$ &$\mp  3$ &$\mp  3$ &$\pm  7$ &$\pm  7$ &$\pm 4$ &$\pm 2.456$ &$\mp 4.264$ \\
11\ \ \  &$\pm 11$ &$\pm 1$  &$\pm 4$ &$\pm  7$ &$\pm  7$ &$\mp  3$ &$\mp  3$ &$\pm 4$ &$\pm 2.456$ &$\pm 4.264$ \\
17\ \ \  &$\pm 1$  &$\pm 17$ &$\pm 6$ &$\mp  5$ &$\mp 5$ &$\pm  11$ &$\pm  11$ &$\pm 6$ &$\pm 3.053$ &$\mp 5.488$ \\
17\ \ \  &$\pm 17$ &$\pm 1$  &$\pm 6$ &$\pm 11$ &$\pm  11$ &$\mp 5$ &$\mp 5$ &$\pm 6$ &$\pm 3.053$ &$\pm 5.488$ \\
23\ \ \  &$\pm 1$  &$\pm 23$ &$\pm 8$ &$\mp  7$ &$\mp 7$ &$\pm  15$ &$\pm  15$ &$\pm 8$ &$\pm 3.551$ &$\mp 6.487$ \\
23\ \ \  &$\pm 23$ &$\pm 1$  &$\pm 8$ &$\pm 15$ &$\pm  15$ &$\mp 7$ &$\mp 7$ &$\pm 8$ &$\pm 3.551$ &$\pm 6.487$ \\
27\ \ \  &$\pm 3$  &$\pm 9$  &$\pm 4$ &$\mp  1$ &$\mp  1$ &$\pm  5$ &$\pm  5$ &$\pm 4$ &$\pm 3.848$ &$\mp 1.633$ \\
27\ \ \  &$\pm 9$  &$\pm 3$  &$\pm 4$ &$\pm  5$ &$\pm  5$ &$\mp  1$ &$\mp  1$ &$\pm 4$ &$\pm 3.848$ &$\pm 1.633$ \\
29\ \ \  &$\pm 1$  &$\pm 29$ &$\pm10$ &$\mp  9$ &$\mp 9$ &$\pm  19$ &$\pm  19$ &$\pm10$ &$\pm 3.988$ &$\mp 7.353$ \\
29\ \ \  &$\pm 29$ &$\pm 1$  &$\pm10$ &$\pm 19$ &$\pm  19$ &$\mp 9$ &$\mp 9$ &$\pm10$ &$\pm 3.988$ &$\pm 7.353$ \\
35\ \ \  &$\pm 1$  &$\pm 35$ &$\pm12$ &$\mp 11$ &$\mp 11$ &$\pm 23$ &$\pm 23$ &$\pm12$ &$\pm 4.381$ &$\mp 8.128$ \\
35\ \ \  &$\pm 5$  &$\pm 7$  &$\pm 4$ &$\pm  1$ &$\pm  1$ &$\pm  3$ &$\pm  3$ &$\pm 4$ &$\pm 4.381$ &$\mp 0.478$ \\
35\ \ \  &$\pm 7$  &$\pm 5$  &$\pm 4$ &$\pm  3$ &$\pm  3$ &$\pm  1$ &$\pm  1$ &$\pm 4$ &$\pm 4.381$ &$\pm 0.478$ \\
35\ \ \  &$\pm 35$ &$\pm 1$  &$\pm12$ &$\pm 23$ &$\pm 23$ &$\mp 11$ &$\mp 11$ &$\pm12$ &$\pm 4.381$ &$\pm 8.128$ \\
\hline
\end{tabular}
\caption{Some values for the action integral, $A(t_0)$, and the relative
interaction strength, $\alpha$, allowed for total population transfer.
These values are found using Eq(\ref{qnumbers'}) subject to the conditions listed.
Here $n_e = n_o + n_o'$, $n_o = \frac{1}{3}(2n_1 - n_2)$ 
and $n_o' = \frac{1}{3}(2n_2 - n_1)$. 
}
\end{table}

Numerical calculations for the time dependence of the populations
in a degenerate $3$-state atom perturbed by external interactions with
$V(t) = V_0 \cos(\omega t)$ are presented in figures~1-4.  
These results were obtained by using a standard fourth order  
Runge-Kutta numerical integration of Eq(\ref{3degen}) with $\epsilon_j =0$
and $\beta = 1$ for various values of $\alpha$ and $A(t_0)$.  For most values of $\alpha$
the population transfer into either level 2 or level 3 is always incomplete.
A typical case is shown in figure~1 where $\alpha = 2$ and $A(t_0) =  1.5$.  
We note that $P_1(t)$ does return to zero twice in each period, $T$,
of the external field, but that complete population transfer to
an initially unoccupied state never occurs.  We also ran calculations (not shown)
for $\alpha = 8.5$ and $A(t_0) = 5.8$, which are not values that lead to complete 
population transfer from Eq(\ref{qnumbers'}).  Again complete transfer to an initially 
unoccupied level never occured.  However, there were rapid oscillations 
in the populations of all states except near $t = T/4$ and $t = 3T/4$,
where none of the populations oscillated rapidly.  This was similar to figure~4.
This appears to correspond to the onset of complete population transfer,
which occurs at $t = T/4$ and odd multiples of $T/4$, as seen in figures~2-4.  

\begin{figure}
\scalebox{0.7}{\includegraphics{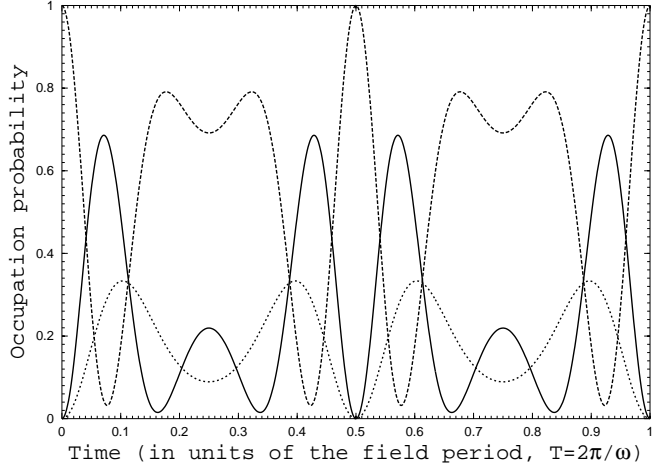}}
\caption{Occupation probabilities as a function of time over one period, $T$,
of the external field, $V(t) = V_0 \cos(\omega t)$.  The solid line is the
probability, $P_2(t)$, that the electron is transferred to the target state,
namely state 2.
The long dash line is the probability, $P_1(t)$, that the electron
is in state 1, the state in which it began initially.  
The short dash line is the probability, $P_3(t)$, that the electron
is in state 3.  In this figure $\alpha = V_{12}/V_{23} = 2$ and the
action area at $t = t_0$ is $A(t_0) = \int_0^{t_0} V(t') dt' = 1.5$.
Since $\alpha$ and $A(t_0)$ do not correspond to values given by Eq(\ref{qnumbers'}), 
complete transfer to level 2 does not occur. 
}
\label{fig1}
\end{figure}

\begin{figure}
\scalebox{0.7}{\includegraphics{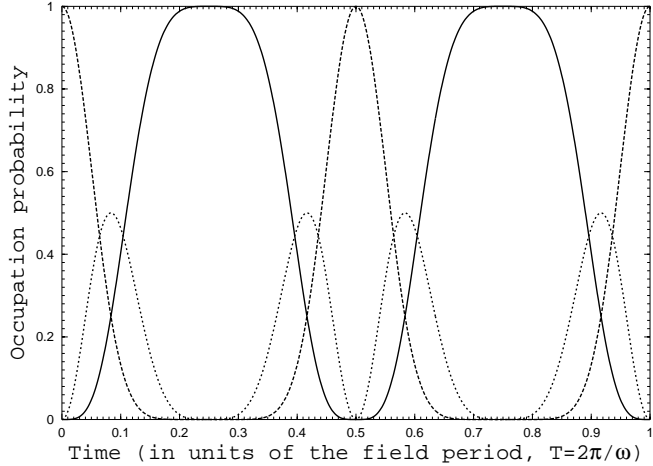}}
\caption{Occupation probabilities as a function of time.
The solid line denotes $P_2(t)$; the long dash line denotes $P_1(t)$;
and the short dash line denotes $P_3(t)$.  
In this figure we use the  allowed values, $\alpha = V_{12}/V_{23} = 0$ and 
$A(t_0) = \int_0^{t_0} V(t') dt' = 2.221$, corresponding
to $n_1 = 3$ and $n_2 = 3$ ($n_o = n_o' = 1$) in Eq(\ref{qnumbers'}).
Complete transfer to level 2 from level 1 occurs at $t = t_0 = T/4$
and again at odd multiples of $t_0$. 
} 
\label{fig2}
\end{figure}

Calculations using a few values of $\alpha$ that permit complete transfer
to level 2 are shown in figures~2-4.  Our numerical codes give the same
results as the analytic expressions of Eq(\ref{Pj'}).  This provides
a check that our algebra is correct.  
We note that $\frac{2 n_1 n_2}{(n_1 + n_2)^2} \leq \frac{1}{2}$ and
that the maximum value of $P_3(t)$ occurs for $n_1 = n_2$, where
$P_{3 \ max} = \frac{1}{2}$, consistent with figure~2.  This corresponds
to $\alpha =0$ so that direct transitions from level 1 to level 2 are
forbidden.  Transfer to level 2 occurs via the intermediate level 3.
Transfer from level 1 to level 2 and back is complete, and
occurs periodically.  
In general $\alpha = 0$ occurs when $n_1 = n_2 = 3n_{odd}$,
where $n_{odd}$ is any odd integer.
This appears to give the simplest condition that allows complete
population transfer.  
In this case the action area is $A(t_0) = n_{odd} \pi / \sqrt{2}$.

\begin{figure}
\scalebox{0.7}{\includegraphics{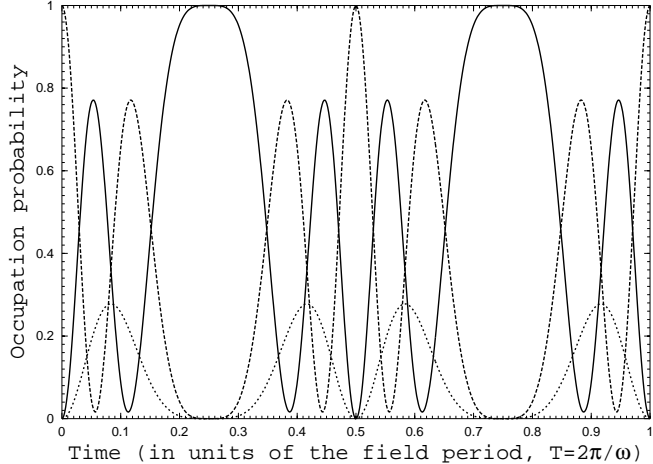}}
\caption{Occupation probabilities as a function of time.
The solid line denotes $P_2(t)$; the long dash line denotes $P_1(t)$;
and short dash line denotes $P_3(t)$.  
In this figure $\alpha = -2.530$ and $A(t_0) = 1.656$, corresponding
to $n_1 = 1$ and $n_2 = 5$ ($n_o= -1$, $n_o' = 3$) in Eq(\ref{qnumbers'}).
}
\label{fig3}
\end{figure}

Calculations for two other values of $\alpha$ that allow complete transfer
to level 2 are shown in figures 3 and 4.  We see that 
complete transfer occurs twice in one period of the oscillating field
but that the frequency of the "side bands" increases as $\alpha A$ increases.  
We note that $\alpha = \pm \sqrt{\frac{2}{n_1 n_2}} \ (n_2 - n_1)$ becomes either large 
($n_2 \gg n_1$, or vice versa), or small ($n_1 \sim n_2$) as $n_1 n_2$ increases,
while $A(t_0)$ increases as $\sqrt{n_2 n_2/2}$.
When complete population transfer occurs, the population lingers in level 2,
as seen in figures~2-4.  
It can be shown \cite{sm02} that near $t = t_0$ $1 - P_2(t)$
varies as $[\omega (t - t_0)]^4$.

\begin{figure}
\scalebox{0.7}{\includegraphics{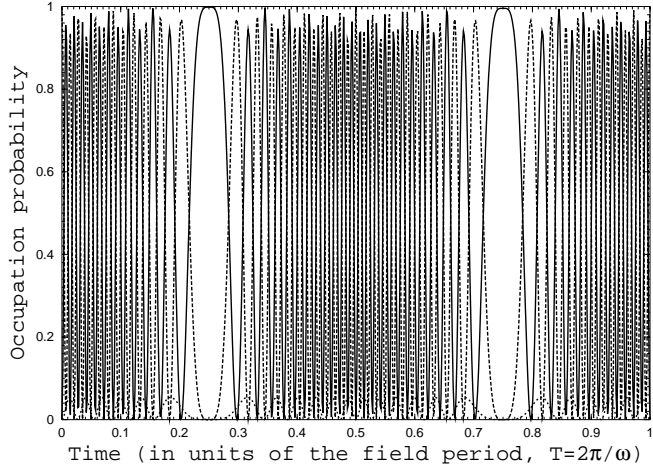}}
\caption{Occupation probabilities as a function of time.
The solid line shows $P_2(t)$; the long dash line shows $P_1(t)$;
and the short dash line shows $P_3(t)$.  
Here $\alpha = 8.128$ and $A(t_0) = 4.381$, corresponding
to $n_1 = 35$ and $n_2 = 1$ ($n_o= 23$, $n_o' = -11$) in Eq(\ref{qnumbers'}).
Since these values are allowed, the electron population
is completely transferred to level 2 from level 1.
}
\label{fig4}
\end{figure}

Additional control \cite{sm02} may be achieved by changing the shape of $V(t)$.
This can be used to control how long the population remains near unity
in level 2, for example.  An interesting example  
is the case of an ideal, sudden 'kick' produced by $V(t) = A_0 \delta(t - t_0)$.  
If $A_0 \to n_{odd} \pi / \sqrt{2}$ for example, then the mere
presence of level 3 allows transfer from level 1 to level 2 
without any direct transfer from level 1 to level 2.
With this ideal 'kick', level 2 is unoccupied before $t = t_0$, 
and fully occupied after $t = t_0$. 

To obtain analytic solutions to Eq.(\ref{ndegen}) for a 3-level atom, 
we imposed degeneracy on the 3-level manifold, e.g. by taking $E_j = E = 0$.
If $E \neq 0$, then $E$ can be removed from Eq(\ref{nexact}) 
by an overall phase transformation, $e^{i E t}$.
When the levels are not degenerate $\omega_{ij} = (E_i - E_j)/\hbar \neq 0$, and
the population transfer is incomplete.
The effect of this population leakage can be calculated by expanding the transition amplitude 
$a_2$ in a power series in time with terms small in $\omega_{ij}t$.
After some algebra we obtain from Eq(\ref{nexact}) the difference between the
degenerate and nearly degenerate populations of level 2, namely, 
$\Delta P_2(t) \simeq \frac{1}{12} [ 2( 2 \omega_{13} 
- \omega_{12}) V_{12}(0)V_{13}(0)V_{23}(0) + \omega_{12}^2 V_{12}^2(0)] t^4$, 
where $V_{ij}(0)$ denotes the value of $V_{ij}(t)$ at $t = 0$.  
By the time the occupation probability $P_2$ reaches its first maximum at 
$t_0 = T/4 = \frac{\pi}{2 \omega}$, using 
$\alpha^{-1} V_{12}(0) = \beta^{-1} V_{13}(0) = V_{23}(0) = V(0) = 
\pm \sqrt{\frac{n_1 n_2}{2}} \frac{\pi}{3} \omega $, 
the difference becomes, 
$ \Delta P_2(t_0) \simeq \frac{1}{27} (\frac{\pi}{2})^6 
[  \frac{\pi}{3} \beta n_1 n_2 (n_2 - n_1) 
( 2 \frac{\omega_{13}}{\omega}- \frac{\omega_{12}}{\omega})
 + (n_2 - n_1)^2 (\frac{\omega_{12}}{\omega})^2 ] $.
The choice of the field frequency, $\omega$, involves a trade-off between the duration 
of time the population remains in state 2 and the population leakage.
The higher the frequency, $\omega$, the smaller the population leakage, $\Delta P_2$,
but the shorter the duration time, $T_s$, that the population remains in level 2.
This effect of population leakage is similar as that for a 2-level atom \cite{sm02}.
Numerical calculations indicate that this population leakage grows rapidly in time.

\section{Discussion}

In this paper we have considered transitions amplitudes and
probabilities in a $3$-level atom with degenerate states.
We also assumed that the interaction matrix elements, $V_{ij}(t)$ 
all have a common time dependence.  In realistic atomic
systems, energy levels are seldom, if ever, exactly degenerate.
Also levels outside $3$-level manifold usually exist.
Consequently, use of our results is restricted to external fields 
with frequencies in the range, $\omega_{min} < \omega <  \omega_{max}$.
Here $\hbar \omega_{min}$ is the energy splitting of the nearly
degenerate states, and $\hbar \omega_{max}$ is the energy difference
between the $3$-level manifold and the closest state in energy outside 
the manifold.   

It is instructive to compare our 3-level results to 
simpler 2-level results \cite{sm02,msr03}.
The equations for a $2$-level atom may be recovered from our $3$-level
equations by taking $V_{12}$ and $V_{13}$ to zero, which corresponds
to $\alpha \to \infty$ (with any $|\epsilon_j/\alpha| < \infty$), 
as may be understood from Eq(\ref{3trial}).  
For the $2$-level atom it has been
shown \cite{msr03,sm02} that $a_j(t)$ may be expressed
in terms of eigenstates $c_i(t)$ via the relation,
$c_i(t) = \sum_{j=1}^2 {\cal M}_{ij} a_j(t)$, where 
${\cal M}_{ij}=\left(\begin{array}{cc}
1&y_+\\
1&y_-\\
\end{array}\right)$.
Here $y_{\pm}$ are the same as those given by Eq(\ref{ypm}).
This matrix may be inverted to give the population amplitudes,
$a_1(t) = \frac{1}{\Delta}[y_- c_1(t) - y_+ c_2(t)] $
and $a_2(t) = \frac{1}{\Delta}[ - c_1(t) +  c_2(t)]$,
where $\Delta = 2 \sqrt{1 + (\frac{\epsilon_2 - \epsilon_1}{2})^2}$
is the determinant of ${\cal M}_{ij}$.
A simple solution occurs when $\epsilon_1 = \epsilon_2 = \epsilon$,
namely,
$ a_1(t) = e^{-i \epsilon A(t)} \ \frac{1}{2} [e^{-iA(t)} + e^{+iA(t)}] 
	=  \ e^{-i \epsilon A(t)} \cos(A(t))$ and 
$ a_2(t) = e^{-i \epsilon A(t)} \ \frac{1}{2} [e^{-iA(t)} - e^{+iA(t)}] 
	= - i e^{-i \epsilon A(t)} \sin(A(t)) $.
Note that $P_1 = 1 - P_2$. 
One may then simply determine the conditions under which the population
of level 2, $P_2(t_0)$, takes on any desired value at time $t_0$.
If $\epsilon_1 \neq \epsilon_2$, it is easily shown that 
$P_2(t_0) \leq 1/(1+ (\epsilon_2 - \epsilon_1)^2/4 ) < 1$. 
Thus the diagonal matrix elements of $V_{ij}(t)$ prevent complete transfer 
to the initially unoccupied level when $\epsilon_1$ and $\epsilon_2$ are 
unequal, and, when they are equal, they simply contribute an overall phase.
When $\epsilon_1 = \epsilon_2$ any value of $P_2(t_0)$ between 0 and 1 can be found. 
In particular, $P_2(t_0)=1$ if 
$A(t_0) = \int_0^{t_0} V_{12}(t') dt' = n_{odd} \pi /2$.
This allowed value of $A(t_0)$ for complete transfer to level 2 differs 
from the allowed values given in Eq(\ref{qnumbers'}) for degenerate 3-level atoms.
By comparison to a 2-level atom the conditions for complete transfer 
are generally more complex in a $3$-level atom, as discussed in the appendix.
For example, complete population transfer can occur in a 3-level atom
when $\epsilon_2 \neq \epsilon_3$.
In both cases, however, the duration of time spent in the transferred state 
can be controlled by adjusting the time dependence of the external field.
Also if the states are not quite degenerate then population leakage occurs.
For a harmonic $V(t)$ the population leakage in a 2-level atom varies as 
$\Delta P_2(t_0) \simeq \frac{1}{4} (\frac{\pi}{2})^6 (\omega_{12}/\omega)^2$. 
For $2s-2p$ transitions in atomic
hydrogen complete population control can be nearly achieved
in this manner using a radiation field with wavelengths (and intensities)
ranging from a few $\mu m$ (with about $10^{12}$ W/cm$^2$) 
up to a few $cm$ (with about $10^4$  W/cm$^2$).
At wavelengths below a few $\mu m$ coupling to nearby $n = 3$ atomic levels
can be significant, and this $2$-level model breaks down. 

While we have not provided calculations in this paper for specific experiments,
some general guidelines for experimental tests and applications are evident.
First, the model must be valid, so that the frequency of the external interaction
is limited by $\omega_{min} < \omega < \omega_{max}$.
If the external interaction varies harmonically, $V(t) = V_0 \cos(\omega t)$,
then the first line of Eq(\ref{qnumbers'}) imposes a constraint between
$V_0$ and $\omega$, namely, $V_0/\omega = \pm \sqrt{\frac{n_1 n_2}{2}} \frac{\pi}{3}$.
For a harmonic interaction the duration of the time, $T_s$, that the population remains
in level 2 varies inversely with $\omega$;  the higher the frequency,
$\omega$, the smaller the time the population remains in level 2.
Our degenerate 3-level model can be applied to systems with dipole
selection rules.  A dipole selection rule can correspond to $\alpha = 0$.
Transfer from level 1 to level 2 (which is the dipole forbidden transition)
is complete at time $t_0$ if the two allowed transition matrix elements,
$V_{23}$ and $V_{13}$, are equal in absolute magnitude and $A(t_0) = n_{odd} \pi/2$.
In some atomic systems state 2 could decay, with a lifetime, $T_d$, to another
level outside the degenerate manifold and be lost.
Such loss can be controlled by adjusting $\omega$.
Our model also rests on degeneracy.
If the three levels are not degenerate the transfer of population to
level 2 is incomplete.  We call this population leakage.
As discussed above this population leakage may be minimized
by using high frequency external interactions. i.e. $\omega \gg \omega_{min}$,
but at a cost to duration of time, $T_s$, the population remains in level 2.

Using nearly degenerate states, the duration time, $T_s$, 
can be further controlled \cite{sm02} by adjusting the shape of $V(t)$.
As mentioned in section III for example, the transfer is complete, instantaneous 
and permanent when $V(t) = n_{odd} \frac{\pi} {\sqrt{2}}  \delta(t - t_0)$.
Such a quick, hard pulse is called \cite{dunning99,burg02,dunning02} 
a 'kick'.  Two practical limitations on this ideal model are 
the impossibility of producing a signal that varies as $\delta(t - t_0)$,
and the existence of an infinitely wide spectrum of high frequency components
with frequencies $\omega > \omega_{max}$ in the Fourier spectrum of $\delta(t - t_0)$.
Fortunately these two difficulties can both be addressed by using
'kicks' of finite width in time.  
In some cases it may be possible to design a 'kick' 
so that its duration is short compared to any other changes in the system,
so that a finite 'kick' may be sensibly represented by $\delta(t - t_0)$.
If, in addition, the energy levels outside the (nearly) degenerate manifold
have a large energy gap $\hbar \omega_{max}$, then it may be possible
that $\omega_{min} < \omega < \omega_{max}$ and our model may be applicable.
Applying $V(t) \simeq \delta(t - t_0)$ to nearly degenerate atomic systems
leads to the 'gedanken' question, what happens when one tries
to force the transition to occur within a small time interval about $t_0$
in a degenerate quantum system where $\Delta t$ is large?

Let us next consider some mathematical features 
that occur when degenerate states are used.
The exact probability amplitude for a transition from
an initial state $|i \rangle$ to a final state $|f \rangle$ may be expressed
in terms of the time evolution operator $U(t_i,t_f)$, namely,
$a_{fi} = \langle f |U(t_f,t_i) |i \rangle $, where (setting $t_i=0$ and $t_f=t$),
$U(t,0) = T e^{-i \int_0^t V(t')dt'} 
= \sum_n  \frac{(-i)^n}{n!}  \int_0^t ... \int_0^t T V(t_n) .... V(t_1) dt_n ... dt_1$.
Here $T$ is the Dyson time ordering operator which places the interactions
in the order of increasing time so that for $t \geq t'$, $T V(t) V(t')$
and $T V(t) V(t') = 0$, and vice versa for $t' \geq t$, whence $T  [V(t), V(t')] \neq 0$.  
In any intermediate time step from $t_j$ to $t_{j+1}$, the matrix element
is $\int_0^t \ e^{(E_I - E_{I'}) t_j} <I|V(t_j)|I'> dt_j$.  
Without the phase $e^{(E_I - E_{I'}) t_j}$ there is no time ordering
since $T [V(t), V(t')] \to 0$ as $T \to 1$.
If all the states are degenerate, then $\Delta E = E_I - E_{I'} \to 0$ in
every time step. 
This corresponds to taking $T \to 1$ in $U(t,0)= T e^{-i \int_0^t V(t')dt'}$.
In this limit of degeneracy, $U(t,0) \to e^{-i\int_0^t V(t') dt'}$,
corresponding to the approximation of Magnus \cite{Magnus}, where $[V(t),V(t')] \to 0$.
In this limit there are no time correlations in the time propagation.
Imposing degeneracy removes time ordering.
Moreover in standard stationary scattering with $H = H_0 + V$, 
the Fourier transform of $ e^{i\int_0^t (E - H)dt}$ is $i \pi \delta(E - H)$,
which is the energy conserving part of the Green function,
$G_0 =  \frac{1}{E - H \pm i \eta} = i \pi \delta(E - H) \mp P_v \frac{1}{E - H }$.
The energy non-conserving quantum fluctuations, $ P_v \frac{1}{E - H } $,
imposed by the asymptotic $\pm i \eta$ initial condition,  
carry the direction of time, i.e., outgoing or incoming scattered waves.
Here $P_v \frac{1}{E - H } $ is the principal value integral that
omits the contribution at $E = H$.
The Fourier transform of $P_v \frac{1}{E - H }$ is ${\rm sign}(t) = T - 1$.
Degeneracy, $P_v \frac{1}{E - H } \to 0$, again removes time ordering.  
Thus, imposing degeneracy mathematically corresponds to, 
\begin{eqnarray}
\label{features}
 \Delta E \to 0 &\Longleftrightarrow& \int_0^t \ e^{(E_I - E_{I'}) t_j}  <I|V(t_j)|I'> dt_j
	\to \int_0^t  \ <I|V(t_j)|I'> dt_j \nonumber \\
 &\Longleftrightarrow& T [V(t),V(t')] \to 0
	 \Longleftrightarrow U(t,0) \to e^{-i \int_0^t V(t')dt'} \nonumber \\
 &\Longleftrightarrow& e^{i(E - H_0)t} T e^{-i \int_0^t V(t')dt'} \to e^{i\int_0^t (E - H)dt} 
\nonumber \\
&\Longleftrightarrow& \frac{1}{E - H \pm i \eta} \to \frac{1}{E - H }
	 \Longleftrightarrow P_v \frac{1}{E - H } \to 0 \nonumber \\
& \Longleftrightarrow& T \to 1 \ \ .
\end{eqnarray}
As a consequence of using degenerate states, quantum energy fluctuations in
intermediate states are eliminated.  The Hilbert space is restricted so that 
$\Delta t = \hbar/\Delta E \to \infty$.  
In addition the minimal size of the wavepacket, $\Delta \ell = v \Delta t$,
now extends to infinity, so that the quantum system may not be decoherently
decoupled from its macroscopic environment. 
This effect is evident in $2s-2p$ transitions in hydrogen
caused by the impact of ions, where the range of the interaction
goes to infinity as the levels become degenerate \cite{ms80,mcg79}.  
Within the interval $\Delta t$ time cannot be localized, e.g. reproducibly observed.
Time ordering i.e. causality, is not operative and there is no 'flow of time'.  
In the classical limit as $\Delta t \to 0$,
time can be measurably localized, being defined by two closely spaced
decoherent observable events.  
In this classical limit causality strictly holds.      

We note that at sufficiently high $\omega$ all bound levels in the atomic elements 
become nearly degenerate.
Hence, if one can deal with high energy continuum states,
our approach might be useful in applications involving fourth generation synchrotrons
that produce intense high frequency fields. 
An extension of this method might also used for control transitions in high Rydberg states
\cite{Gallagher}, including adiabatic rapid passage \cite{hk83,gallagher02}.
RWA is used to describe coherent storage of information in photonic states \cite{mair02}.
The approach developed here may be useful for modeling information 
transmission and storage in atomic states \cite{cirac97} in a new way.

\section{Summary}
We have shown that in a 3-level atom with degenerate energies, electron population
is completely transferred via an external interaction at a designated time, $t_0$,
from an initially occupied level (level 1) to a designated initially unoccupied
level (level 2) under two conditions.  
The first condition for complete transfer is that the ratio of the matrix elements 
of the external interaction, $V_{ij}(t)$,satisfy $V_{12}(t)/V_{23}(t) 
= \alpha = \pm \sqrt{\frac{2}{n_1 n_2}} (n_2 - n_1)$,
and $V_{13}(t)/V_{23}(t) = \beta = \pm 1$.
The second condition is that at $t = t_0$ the action area of $V(t)$ satisfy
$A(t_0) = \int_0^{t_0} V(t') dt' = \pm \sqrt{\frac{n_1 n_2}{2}} \frac{\pi}{3}$,
where we have set $V_{23}(t) = V(t)$.
Here $n_1$ and $n_2$ are integers such that $n_1 = 2 n_o + n_o'$ and
$n_2 = n_o + 2n_o'$, where $n_o$ and $n_o'$ are any odd integers.
The duration of time the transferred population remains in level 2
can be controlled either by varying the frequency, $\omega$, of the 
external potential, or by varying the shape of $V(t)$.

\begin{acknowledgments}
We thank J.H. Eberly, B.W. Shore and A.L. Godunov for useful discussion.
This work was supported in part by the Division of Chemical Sciences, Office
of Sciences, U.S. Department of Energy.  KhR is supported by a NSF-NATO Fellowship.
\end{acknowledgments}

\appendix

\section{Method to find conditions for complete population transfer}

Here we show how to find the conditions on the $V_{jk}$ such that the electron population
is completely transferred from its initial state $i = 1$ to a different
final state, $i = 2$ or $3$.  A convenient way to begin is to take
$d P_k(t)/dt = 0$ at $t = t_0$.  This gives both maxima and minima for $P_k$. 
Using Eq(\ref{Pj}) this yields,
\begin{eqnarray}
\sin((z_i-z_j)A(t_0)) &=& 0   \\ \nonumber
(z_i - z_j)A(t_0) &=& m_{ij} \pi \ \   (m_{ij} \neq 0) \ \  (i,j=1,2,3) \ \ .
\end{eqnarray}
Using $\cos((z_i-z_j)A(t_0))=\cos(m_{ij}\pi)=(-1)^{m_{ij}}$, one has,
\begin{eqnarray}
&P_1=|a_1|^2=\frac{1}{\Delta^2}[(x_2y_3-x_3y_2)^2+(x_3y_1-x_1y_3)^2+(x_1y_2-x_2y_1)^2 \nonumber \\
&+2(-1)^{m_{12}}(x_2y_3-x_3y_2)(x_3y_1-x_1y_3) \nonumber \\
&+2(-1)^{m_{13}}(x_2y_3-x_3y_2)(x_1y_2-x_2y_1)\nonumber \\
&+2(-1)^{m_{23}}(x_3y_1-x_1y_3)(x_1y_2-x_2y_1)]\nonumber \\
&=\frac{1}{\Delta^2}[(-1)^{k_1}(x_2y_3-x_3y_2)+(-1)^{k_2}(x_3y_1-x_1y_3)+(-1)^{k_3}(x_1y_2-x_2y_1)]^2 \ \ ,
\end{eqnarray}
where $ k_1+k_2=m_{12}$, $k_1+k_3=m_{13}$ and $ k_2+k_3=m_{23}$.
Now,
\begin{eqnarray}
k &\equiv& k_1 = \frac{1}2(m_{12}+m_{13}-m_{23})
=\frac{1}2(\frac{A(t_0)}{\pi}(z_1-z_2)+\frac{A(t_0)}{\pi}(z_1-z_3)-\frac{A(t_0)}{\pi}(z_2-z_3))
\nonumber \\ \nonumber 
  && \ \ \ \ = \frac{A(t_0)}{\pi}(z_1-z_2) \\ \nonumber 
&&  k_2 = \frac{1}2(m_{12}+m_{23}-m_{13})
=\frac{1}2(\frac{A(t_0)}{\pi}(z_1-z_2)+\frac{A(t_0)}{\pi}(z_2-z_3)-\frac{A(t_0)}{\pi}(z_1-z_3))=0
\\ \nonumber 
k' &\equiv& k_3 =\frac{1}2(m_{13}+m_{23}-m_{12})
=\frac{1}2(\frac{A(t_0)}{\pi}(z_1-z_3)+\frac{A(t_0)}{\pi}(z_2-z_3)-\frac{A(t_0)}{\pi}(z_1-z_2))
\nonumber \\ 
  && \ \ \ \ = \frac{A(t_0)}{\pi}(z_2-z_3)
\end{eqnarray}
This yields extrema for $P_k$ at $t = t_0$, namely,
\begin{eqnarray}
P_1&=&\frac{1}{\Delta^2}[(-1)^k(x_2y_3-x_3y_2)+(x_3y_1-x_1y_3)+(-1)^{k'}(x_1y_2-x_2y_1)]^2 
\nonumber \\
P_2&=&\frac{1}{\Delta^2}[(-1)^k(y_2-y_3)+(y_3-y_1)+(-1)^{k'}(y_1-y_2)]^2 \nonumber \\
P_3&=&\frac{1}{\Delta^2}[(-1)^k(x_3-x_2)+(x_1-x_3)+(-1)^{k'}(x_2-x_1)]^2 
\end{eqnarray}
where $k$ and $k'$ are non zero integers that are {\it not both even}.

\subsubsection{Conditions for complete transfer}

We now seek conditions such that at $t = t_0$, $P_1=0$, $P_2=1$, and $P_3=0$.
There are three sets, $\{k,k'\}$, that satisfy these conditions,
namely, i) $k$ is even and $k'$ is odd, ii) $k$ and $k'$ are both odd, 
and iii) $k$ is odd and $k'$ is even.

\underline{\it Case i}: $k$ is even and $k'$ is odd.  Then, 
\begin{eqnarray}
P_1&=&\frac{1}{\Delta^2}[(x_2y_3-x_3y_2)+(x_3y_1-x_1y_3)-(x_1y_2-x_2y_1)]^2 \nonumber \\
P_2&=&\frac{1}{\Delta^2}[(y_2-y_3)+(y_3-y_1)-(y_1-y_2)]^2=\frac{4}{\Delta^2}(y_2-y_1)^2 
\nonumber \\
P_3&=&\frac{1}{\Delta^2}[(x_3-x_2)+(x_1-x_3)-(x_2-x_1)]^2=\frac{4}{\Delta^2}(x_1-x_2)^2  \ \ .
\end{eqnarray}
From the condition that $P_2=1$ one has $[2(y_2-y_1)/\Delta]^2=1$, 
where $y_1\not=y_2$, so that $\Delta=\pm 2 (y_2-y_1)$.
The condition that $P_3=0$ gives $[2(x_1-x_2)/\Delta]^2=0$, which yields $x_1=x_2=x$.
Using this result one then obtains $P_1=\left[(x+x_3)(y_1-y_2)/\Delta\right]^2$.
Requiring $P_1=0$, one has $x_3=-x$.
Now, using $x_1=x_2=-x_3=x$, we have from below Eq(\ref{Minv}), $\Delta=2x(y_2-y_1)$.
Finally since $\Delta=\pm 2 (y_2-y_1)$, one has $x_1=x_2=-x_3=x=\pm 1$.

\underline{\it Case ii}: $k$ and $k'$ are both odd.  Then,
\begin{equation}
\begin{array}{l}
P_1=\frac{1}{\Delta^2}[-(x_2y_3-x_3y_2)+(x_3y_1-x_1y_3)-(x_1y_2-x_2y_1)]^2 \\
P_2=\frac{1}{\Delta^2}[-(y_2-y_3)+(y_3-y_1)-(y_1-y_2)]^2=\frac{4}{\Delta^2}(y_3-y_1)^2 \\
P_3=\frac{1}{\Delta^2}[-(x_3-x_2)+(x_1-x_3)-(x_2-x_1)]^2=\frac{4}{\Delta^2}(x_1-x_3)^2 \,. \\
\end{array}
\end{equation}
From the condition that $P_2=1$ one has $[2(y_3-y_1)/\Delta]^2=1$, 
where $y_1\not=y_3$, so that $\Delta=\pm 2 (y_3-y_1)$.
The condition that $P_3=0$ gives $[2(x_1-x_3)/\Delta]^2=0$, 
which yields $x_1=x_3=x$. Then one obtains 
$P_1=\left[(x+x_2)(y_1-y_3)/\Delta\right]^2$.
When $P_1=0$ one has $x_2=-x$.
Now, using $x_1=-x_2=x_3=x$, we get from below Eq(\ref{Minv}), $\Delta=2x(y_1-y_3)$.
Since $\Delta=\pm 2 (y_3-y_1)$, one has $x_1=-x_2=x_3=x=\mp 1$.

\underline{\it Case iii}: $k$ is odd and $k'$ is even.  This is similar
to case i).  One obtains, $ \Delta=2x(y_3-y_2) $ and 
$-x_1=x_2=x_3=x=\pm 1$.

Complete population transfer, i.e. $P_2 = 1$, requires that $x=\pm1$, 
so that the relative interaction strengths $\alpha$ and $\beta$ must satisfy,
\begin{eqnarray}
\label{ab}
\alpha&=&\frac{2}y-y\pm\frac{\epsilon_2-\epsilon_1}{y^2}\pm(\epsilon_3-\epsilon_2) 
\nonumber \\
\beta&=&\pm 1 +\frac{\epsilon_2-\epsilon_1}{y}
\end{eqnarray}
Next we use these results to find the conditions on $V_{ij}$ required
for complete population transfer to occur.

\subsubsection{Quantum numbers for $A(t_0)$ and $\alpha$}

We now note that relatively simple conditions occur when 
$\epsilon_1=\epsilon_2=\epsilon_3=\epsilon$, namely,
$\alpha=\frac{2}y-y$ and $\beta=\pm 1$.
In this case Eq(\ref{ab}) reduces from a cubic to an easier
quadratic equation in $y$.  This quadratic equation corresponds
to three roots of the more general cubic equation, namely,
\begin{eqnarray}
\label{ypm}
  \{y_j\} = \{y_+, y_-, 0 \}  
  = \{\frac{-\alpha + \sqrt{\alpha^2+8}}{2}, \frac{-\alpha - \sqrt{\alpha^2+8}}{2}, 0 \} \ \ . 
\end{eqnarray}
 
Now for \underline{case i)} when $k$ is even and $k'$ is odd, one has from above,
$x=\{\pm 1,\pm 1,\mp 1\}$ and $y=\{y_+,y_-,0\}$.
From Eq(\ref{ypm}),
$z= \{\mp y_-, \mp y_+, \mp\alpha \}$.
Then, with $E = A(t_0)/\pi$, the useful integers $k$ and $k'$ are: 
$k=\frac{A(t_0)}{\pi}(z_1-z_2)=\pm E\sqrt{\alpha^2+8}$, and 
$k'=\frac{A(t_0)}{\pi}(z_2-z_3)=\pm \frac{E}2(3\alpha-\sqrt{\alpha^2+8})$.
After a little algebra one obtains $(k-k')(2k+k')=18E^2 = n_1 n_2$.
For \underline{case ii)} when $k$ and $k'$ are both odd,
$x=\{\pm 1,\mp 1,\pm 1\}$ and $y=\{y_+,0,y_-\}$.
and $z=\{ \mp y_-, \mp \alpha, \mp y_+ \}$.
Then,
$k=\frac{A(t_0)}{\pi}(z_1-z_2)=\pm \frac{E}{2}(3\alpha+\sqrt{\alpha^2+8})$,
and
$k'=\frac{A(t_0)}{\pi}(z_2-z_3)=\mp \frac{E}2(3\alpha-\sqrt{\alpha^2+8})$.
One may now verify that $(2k+k')(k+2k')=18E^2 = n_1 n_2$.
For \underline{case iii)} when $k$ is odd and $k'$ is even,
$x=\{\mp 1,\pm 1,\pm 1\}$ and $y=\{0,y_+,y_-\}$,
and $z= \{ \mp \alpha, \mp y_-, \mp y_+ \}$.
Then,
$k=\frac{A(t_0)}{\pi}(z_1-z_2)=\mp \frac{E}{2}(3\alpha+\sqrt{\alpha^2+8})$,
and
$k'=\frac{A(t_0)}{\pi}(z_2-z_3)=\pm E\sqrt{\alpha^2+8})$.
One may again verify that $(2k'+k)(k'-k)=18E^2 = n_1 n_2$.

Using $\alpha^2 = k^2/E^2 - 8$ one readily finds in \underline{case i)},
where $k$ is even and $k'$ is odd, that 
$k = \frac{1}{3}(n_1 + n_2)$ and $k' = \frac{1}{3}(n_2 - 2n_1)$.
In \underline{case ii)}, where $k$ and $k'$ are both odd, one
has $k = \frac{1}{3}(2n_1 - n_2)$ and $k' = \frac{1}{3}(2n_2 - n_1)$.  
In \underline{case iii)} 
$k = \frac{1}{3}(n_1 - 2n_2)$ and $k' = \frac{1}{3}(n_1 + n_2)$.
In all three cases one readily obtains the same equation for the
allowed values of $A(t_0)$ and $\alpha$, namely,
\begin{eqnarray}
\label{qnumbers}
   A(t_0)&=& \pm \sqrt{\frac{n_1n_2}{2}} \ \ \frac{\pi}{3} \nonumber \\ 
 \alpha&=& \pm \sqrt{\frac{2}{n_1 n_2}} \ \ (n_2 - n_1) \ \ . 
\end{eqnarray}
It is remarkable that all three cases give the same formula for complete population transfer. 
The top equation corresponds to a constraint on the value of the action,
$\int_0^t V(t') dt'$, i.e. the "area" of the external interaction, $V(t)$, 
or the phase shift caused by the external interaction.  
The bottom equation above gives the relative 
values of the $V_{ij}$.

\section{General solutions for the degenerate 3-level atom}

In the main text a solution for the degenerate 3-level atom was
found in the case when $\epsilon_j = 0$.
We now choose $ \epsilon_j^{-1} V_{jj}(t) = V(t) = V_{23}(t) 
= \alpha^{-1} V_{12}(t) = \beta^{-1} V_{13}(t)$.  
Then Eq(\ref{ndegen}) becomes,
\begin{eqnarray}
\label{3degen'}
i \dot{a}_1(t) &=&  \epsilon_1 V(t) a_1(t) + \alpha V(t) a_2(t) + \beta V(t) a_3(t) \nonumber \\
i \dot{a}_2(t) &=&  \alpha V(t) a_1(t) + \epsilon_2 V(t) a_2(t) + V(t) a_3(t) \nonumber \\
i \dot{a}_3(t) &=&  \beta V(t) a_1(t)  + V(t) a_2(t) + \epsilon_3 V(t) a_3(t)  \ \ .
\end{eqnarray}

Consider a linear superposition $c(t) =  a_1(t) + x \ a_2(t) + y \ a_3(t)$,
where $x$ and $y$ are some time-independent coefficients that we next determine.
Then,
\begin{eqnarray}
\label{3trial'}
i \dot{c}(t) &=& i \dot{a}_1(t) + x \ i \dot{a}_2(t) + y \ i \dot{a}_3(t) \nonumber \\ 
 &=&  \  \epsilon_1 V(t) \ a_1(t) + \alpha V(t) \ a_2(t) + \beta V(t) \ a_3(t) \nonumber \\ 
	&& + x [\alpha V(t) \ a_1(t) + \epsilon_2 V(t) \ a_2(t) + V(t) \ a_3(t)] \nonumber \\ 
	&& + y [\beta V(t) \ a_1(t) + V(t) \ a_2(t) + \epsilon_3 V(t) \ a_3(t)]  \ \ .
\end{eqnarray}
Set $z = \epsilon_1 + \alpha x + \beta y$.  Then,
$i \dot{c}(t) = z V(t)[ a_1(t) +   
 (\frac{\alpha + \epsilon_2 x + y}{\epsilon_1 + \alpha x + \beta y}) a_2(t)
+ (\frac{\beta + x + \epsilon_3 y}{\epsilon_1 + \alpha x + \beta y}) a_3(t) ]$.
We require that $ i \dot{c}(t) =  z V(t) c(t)$.
This holds if and only if $z = \epsilon_1 +  \alpha x + \beta y$,  
$ x = (\alpha + \epsilon_2 x + y)/(\epsilon_1 + \alpha x + \beta y)$ 
and $y = (\beta + x + \epsilon_3 y)/(\epsilon_1 + \alpha x + \beta y)$.
After some algebra this leads to the useful cubic equation,
\begin{eqnarray}
\label{cubicy'}  
 &&[(\beta^2 -\alpha^2) + \alpha\beta(\epsilon_2-\epsilon_3)]y^3
+[\alpha(2-\alpha^2-\beta^2)+\beta(2\epsilon_1-\epsilon_2-\epsilon_3)
+\alpha(\epsilon_1-\epsilon_3)(\epsilon_2-\epsilon_3)]y^2  \nonumber \\
&&+[(2\alpha^2-\beta^2-1) + \alpha\beta(2\epsilon_3-\epsilon_1-\epsilon_2)
+(\epsilon_1-\epsilon_2)(\epsilon_1-\epsilon_3)]y
+[\alpha(\beta^2-1)-\beta(\epsilon_1-\epsilon_2)] \nonumber \\
 &&= 0  \ \ .
\end{eqnarray}
This determines three sets of eigenvalues, $\{x_j\}$, $\{y_j\}$ and $\{z_j\}$,  
and three eigenfunctions, $ c_j(t) = e^{-i z_j  A(t)}$, where 
$A(t) = \int_0^t V(t')dt'$.

When $\epsilon_j \neq 0$ complete population transfer can also occur.
Using $\epsilon_j  = \epsilon \neq 0$ adds nothing new, 
since it produces only an overall phase $e^{i\epsilon t}$. 
A slightly more general, but less elegant, solution for the degenerate
$3$-level equations is found when $\epsilon_1 = \epsilon_2 \neq \epsilon_3$.
Again in this case $\alpha=\frac{2}y-y$ and $\beta=\pm 1$.
The quadratic equation for $y$ now becomes 
$y^2 + \frac{\alpha (1 - \alpha^2) + \alpha(\epsilon_1 - \epsilon_3)^2 
\pm (\epsilon_1 - \epsilon_3)}
{1 - \alpha^2 \pm \alpha(\epsilon_1 - \epsilon_3)} y - 2 = 0$.  
Taking $\tilde{\alpha} = \frac{\alpha (1 - \alpha^2) 
+ \alpha(\epsilon_1 - \epsilon_3)^2 \pm (\epsilon_1 - \epsilon_3) }
{1 - \alpha^2 \pm \alpha(\epsilon_1 - \epsilon_3)}$
the algebra is the same as that one above, except that $\alpha \to \tilde{\alpha}$.
This includes an additional parameter, $ (\epsilon_1 - \epsilon_3)$,
gives simple allowed values of $\tilde{\alpha}$,
and imposes a constraint between $\alpha$ and $(\epsilon_1 - \epsilon_3)$.
More complex conditions for complete population transfer might exist
when $\epsilon_1 \neq \epsilon_2$, in which case Eq(\ref{cubicy'}) is cubic in $y$.
Finally it might be possible that solutions exist when the matrix elements
$V_{ij}(t)$ have different time dependences, i.e. when $\alpha$ and $\beta$
depend on $t$.

Conditions for complete transfer into level 3 at $t = t_0$
may be generally found similarly by interchange of the indices 2 and 3, 
with corresponding interchange of $\alpha$ with $\beta$ and $x$ with $y$.

\end{document}